\documentclass[aps,prl,twocolumn,groupedaddress,nofootinbib,superscriptaddress,notitlepage,showpacs,floatfix]{revtex4-1}
\usepackage{graphicx,graphics}
\usepackage{epsfig}
\usepackage{subfigure}
\usepackage{lipsum}
\usepackage{dcolumn}   
\usepackage{array}
\usepackage{amsthm}
\usepackage{amsmath}
\usepackage{amssymb}
\usepackage{amsfonts}
\usepackage{mathrsfs} 
\usepackage{epsfig}
\usepackage{bm}
\usepackage{braket} 
\usepackage{enumitem}
\usepackage{url}
\usepackage[pdfstartview=FitH]{hyperref}
\usepackage{pifont}
\hypersetup{
    colorlinks=true,        
    linkcolor=blue,              
    citecolor=blue,        
    filecolor=blue,          
    urlcolor=blue,              
    runcolor=cyan
			}
\usepackage[pdftex]{color}
\usepackage{multirow}
\usepackage{tikz}
\usepackage{CJK}
\newcommand{\defeq}{:=}           
\newcommand{\tr}{\mathrm{tr}}
\newcommand{\ud}{\mathrm{d}}

\newcommand{\Saverage}[1]{\Braket{#1}_\mathrm{av}}
\newcommand{\est}{\mathrm{est}}

\begin{document}
\begin{CJK*}{GB}{}
\title{Information Theoretical Analysis of Quantum Metrology}
\author{Yi Peng}
	\affiliation{Institute of Physics, Chinese Academy of Sciences, Beijing 100190, China}
	\affiliation{School of Physical Sciences, University of Chinese Academy of Sciences, Beijing 100190, China}
\author{Heng Fan}
	\email{hfan@iphy.ac.cn}
	\affiliation{Institute of Physics, Chinese Academy of Sciences, Beijing 100190, China}
	\affiliation{CAS Center for Excellence in TQC, University of Chinese Academy of Sciences, Beijing 100049, China}
	\affiliation{Collaborative Innovation Center of Quantum Matter, Beijing 100190, China}
\date{\today}
\eid{identifier}
\pacs{}
\begin{abstract}
	We address the framework of analysing quantum metrology in the information-theoretic picture. 
	Firstly we show how to extract the maximum amount of information in general via suitable state initialization of the probes at 
	the beginning and a quantum measurement at the end. Our analysis can apply to both the single-parameter and the multi-parameter 
	estimation procedures as well as to many other quantum information processing procedures. We then establish 
	a direct connection between the information-theoretic picture of quantum metrology and its conventional variance-covariance 
	picture, by showing that any estimation procedure achieves Heisenberg limit in variance-covariance picture can also reach the 
	information-theoretic Heisenberg limit in the asymptotic sense. As a direct consequence, we argue that the entangled measurement 
	is not necessary for achieving Heisenberg limit in the information-theoretic pictures, which is  explicitly illustrated for the 
	Quantum-Classical Parallel strategy of quantum metrology with a separable measurement employed and the Heisenberg limit saturated
	in both pictures.
\end{abstract}
\maketitle
\end{CJK*}
\emph{Introduction.}---
	The introduction of the concept of information to physics can trace far back to the era of Maxwell. Its intimate relation with
	thermodynamics has been greatly appreciated since~\cite{Maxwell1902Theory,Szilard1929, brillouin1962science,
	landauer1961irreversibility,Penrose1970Foundations, maruyma2009colloquium,Jaynesb1957,Jaynes1957b, Levitin1987,kafri2012holevo,
	Brandao2013a,reeb2014improved,gour2015resource}. It was until the seminal work of Shannon that the framework of information 
	theory had been systematically developed and established for classical information processing 
	devices~\cite{shannon2001mathematical,cover2006elements}, after which the potential of information processing in quantum systems 
	has been realized and thus begins the study of quantum information~\cite{bennett1984quantum,Levitin1987,Bennett1992Communication,
	Bennett1993Teleportation,nielsen2010quantum}. The information-theoretic methodology has hence been widely 
	employed in the study of many quantum technologies such as quantum optimal control~\cite{Lloyd2014}, quantum error 
	correction~\cite{Nielsen1997}, quantum state discrimination~\cite{Holevo1973,Yuen1993,Wootters1973,Davies1978information,Jozsa1994} and even 
	in the fundamental research of quantum mechanics~\cite{Deutsch1983,Levitin1987,maassen1988general,caves1996quantum,
	berta2010uncertainty, Buscemi2014, Coles2015}. Though the temptation to built a information-theoretic picture of quantum metrology
	has always been common~\cite{Ban1997,Sasaki1999,Chiribella2006,Bahder2011,Hall2012Heisenberg-style}, it was addressed 
	directly only recently~\cite{Hall2012,Hassani2017,Hall2018}. The popularity of information-theoretic methodology in physics is partly due
	to the intriguing insight it provides to the study of thermodynamics~\cite{Maxwell1902Theory,Szilard1929, brillouin1962science,
	landauer1961irreversibility, Penrose1970Foundations, maruyma2009colloquium,Jaynesb1957,Jaynes1957b, Levitin1987,kafri2012holevo,
	Brandao2013a,reeb2014improved, gour2015resource}. It is natural to hope to gain similar advantage in the information-theoretic
	framework in other fields. Besides, an unified information-theoretic framework can provides tools to 
	interdisciplinary researches. For instance, it could be beneficial in the study of the thermodynamic effects in aforementioned 
	quantum technologies which is a very important issue in the development quantum technologies~\cite{Lipka-Bartosik2018}.

\emph{Information-theoretic framework of quantum metrology.}---
	As a study of utilizing quantum effects to achieve better estimation of parameters of a physical system, quantum metrology has been
	among the most vigorous branches of quantum technology~\cite{giovannetti2004quantum-enhanced,giovannetti2011advances,Degen2017}. 
	It has great applications in phase estimation~\cite{Mitchell2004,VanDam2007,Liu2015,Demkowicz-Dobrzanski2015}, clock 
	synchronization~\cite{Giovannetti2001,DeBurgh2005,Zhang2013,Yao2017,Zhang2016}, gravitation wave 
	detection~\cite{Buonanno2001,Schnabel2010,Flaminio2017}, imaging of biological samples~\cite{taylor2015quantum,Taylor2016} and many
	other areas~\cite{giovannetti2006quantum,giovannetti2011advances}. Generally, a metrology task is to estimate a parameter or 
	parameters  $\pmb{\pmb{\varphi}}$ using some chosen physical systems as probes. The information of $\pmb{\varphi}$ would be branded
	on the probes after sending them through the quantum channel characterized by $\pmb{\varphi}$. Assume that there are $n$ parameters 
	to be estimated each of which is an entry of the vector $\pmb{\varphi}$. In the noiseless case, it is assumed that the information 
	of $\pmb{\varphi}$ can be imprinted on a probe via the  unitary process 
	$\hat{U}_{\pmb{\varphi}}=\exp(-i\sum_{i=1}^n\varphi_i\hat{\mathcal{H}}_i)$ where $\varphi_i$ denotes the $i$th entry of $\pmb{\varphi}$
	and every Hermitian operators $\hat{\mathcal{H}}_i$ is within our prior knowledge. One can either employ $N$ such probes and
	implement one parameter-imprinting procedure $\hat{U}_{\pmb{\varphi}}$ on each of the probes (parallel strategies) or employ a main probe
	on which the experimenter deploy $\hat{U}_{\pmb{\varphi}}$ repeatedly for $N$ times (sequential strategies). The parameter-imprinting 
	task would be assisted with some carefully designed quantum gates and even some auxiliary probes in the cases of sequential strategies.
	At the end of the evolution, we deploy a general POVM measurement on the probes to generate the experiment data.
	From the information theory point of view, the task of quantum metrology is to extract the maximum amount of information about the
	parameters to be estimated which is evaluated by the mutual information between the measurement outcome $m$ and 
	$\pmb{\varphi}$~\cite{Holevo1973,Yuen1993,Wootters1973,Davies1978information,Jozsa1994,Hassani2017}
	\begin{equation}
		H(m:\pmb{\varphi}) = H(m) + H(\pmb{\varphi}) - H(m,\pmb{\varphi}),
	\end{equation}
	where $H(m)$ and $H(\pmb{\varphi})$ are Shannon information corresponding to the measurement outcome and the parameters while
	$H(m,\pmb{\varphi})$ is the joint Shannon information of them. The search for strategies for the maximum information extraction
	has been addressed by many. While useful bounds has been found and widely 
	used~\cite{Holevo1973,Yuen1993,Wootters1973,Jozsa1994,Hassani2017},
	it has been solved for few specific cases~\cite{Davies1978information,Levitin1987,Ban1997,Sasaki1999}. The development of a general
	formulation for nailing down the optimal information extraction strategy is not only essential for the establishment of the
	information-theoretic framework, but also very import by itself.

\emph{Extraction of the maximum amount of information in quantum metrology.}---
	To achieve the maximum information extraction of the parameters $\pmb{\varphi}$, our task is to choose the optimal combination of the
	state initialization $\hat{\rho}_0$ of the probes before applying the quantum channel characterized by $\pmb{\varphi}$, and the
	quantum measurement at the end. The measurement can be the most general POVM measurement described by $M$ positive 
	Hermitian operators $\hat{\pi}_m$ satisfying the completeness relation $\sum_{m=1}^M\hat{\pi}_m=\openone$.
	In the case of parallel strategy where $N$ probes are employed simultaneously, the probes are initialized to 
	$\hat{\rho}_0$, pass through the parameter channel $\hat{U}_{\pmb{\varphi}}^{\otimes{N}}$ and arrive at the output state 
	$\hat{\rho}_{\pmb{\varphi}}=\hat{U}_{\pmb{\varphi}}^{\otimes{N}}\hat{\rho}_0(\hat{U}_{\pmb{\varphi}}^{\otimes{N}})^\dag$.
	Given $\pmb{\varphi}$, the probability of obtaining the $m$th outcome is~\cite{nielsen2010quantum}
	\begin{equation}
		p(m|\pmb{\varphi})
		=\tr(\hat{\pi}_m\hat{\rho}_{\pmb{\varphi}}).
		\label{p_m_cond_varphi}
	\end{equation}
	Let the prior probability of $\pmb{\varphi}$ be $q_{\pmb{\varphi}}$. The probability of obtaining $m$ on average is 
	\begin{equation}
		p_m = \int\ud^n\pmb{\varphi}q_{\pmb{\varphi}}p(m|\pmb{\varphi}). 
		\label{prob_m}
	\end{equation}
	It turns out that we can always find a strategy where we prepare probes in pure state 
	$\hat{\rho}_0=\ket{\psi_0}\bra{\psi_0}$ and perform the rank-one indecomposable POVM  measurement
	$\hat{\pi}_m=\lambda_m\ket{u_m}\bra{u_m}$ with $\hat{\pi}_m$s being linearly independent, to ensure the maximum information
	extraction~\cite{Davies1978information,PARTHASARATHY1999,DAriano2005}. We remark that the POVM measurement is indecomposable
	if it can not be implemented by deploying other POVM measurements according to a probability 
	distribution~\cite{PARTHASARATHY1999,DAriano2005}. And the necessary conditions for the optimal 
	information extraction are 
	\begin{equation} 
		\int\ud^{n}\pmb{\varphi}p_{\pmb{\varphi}} 
		\braket{u_{m}|\hat{\rho}_{\pmb{\varphi}}|u_{m'}} 
		\ln\left\lbrack{\frac{p_mp(m'|\pmb{\varphi})}{p_{m'}p(m|\pmb{\varphi})}}\right\rbrack 
		=0 
		\label{maxInfoExtrac_nC_r1POVM} 
	\end{equation} 
	for $m\neq{m'}$ and $m,m'=1,\ldots,M$, and 
	\begin{equation} 
		\bra{\psi_0^\perp}
		\int{\ud^{n}\pmb{\varphi}}p_{\pmb{\varphi}}\sum_{m=1}^M\ln\left\lbrack{\frac{p_{{m}}}{p({m}|\pmb{\varphi})} 
		}\right\rbrack 
		(\hat{U}_{\pmb{\varphi}}^{\otimes{N}})^\dag\hat{\pi}_m 
		\hat{U}_{\pmb{\varphi}}^{\otimes{N}}\ket{\psi_0} 
		=0 
		\label{maxInfoExtrac_nC_r1_pureInitial}
	\end{equation} 
	for any $\ket{\psi_0^\perp}$ orthogonal to $\ket{\psi_0}$.
	
	Before moving to the derivation of (\ref{maxInfoExtrac_nC_r1POVM}) and (\ref{maxInfoExtrac_nC_r1_pureInitial}), 
	we have a few remarks on the necessary conditions (\ref{maxInfoExtrac_nC_r1POVM}) and 
	(\ref{maxInfoExtrac_nC_r1_pureInitial}). First of all, they are conditions cover all the optimal information extraction 
	strategies which employ pure initial state and rank-one indecomposable POVM measurement. Secondly, 
	(\ref{maxInfoExtrac_nC_r1POVM}) and (\ref{maxInfoExtrac_nC_r1_pureInitial}) can not rule
	out the existence of other types of optimal information extraction strategies which may utilize mixed initial state or 
	measurement which is not rank-one indecomposable. If there is any such strategy, it can be realized by 
	deploying the optimal strategies given by (\ref{maxInfoExtrac_nC_r1POVM}) and (\ref{maxInfoExtrac_nC_r1_pureInitial}). 
	This is due to the fact the mutual information $H(m:\pmb{\varphi})$ is convex in terms of the initial state and the POVM
	measurement employed. Thus, with (\ref{maxInfoExtrac_nC_r1POVM})  and (\ref{maxInfoExtrac_nC_r1_pureInitial}) we can 
	construct all the optimal information extraction strategies.  Thirdly, every optimal rank-one indecomposable POVM 
	measurement is completely orthogonal, namely $\hat{\pi}_m\hat{\pi}_{m'}=\hat{\pi}_m\delta_{mm'}$~\cite{Jozsa1994}, when 
	the constraints given by (\ref{maxInfoExtrac_nC_r1POVM}) and (\ref{maxInfoExtrac_nC_r1_pureInitial}) are independent. In
	such circumstances, the degrees of freedom of the strategic combination of pure state and rank-one indecomposable POVM 
	measurement with $M$ outcomes should be $-(M-d^N)^2$. Thus, (\ref{maxInfoExtrac_nC_r1POVM}) and 
	(\ref{maxInfoExtrac_nC_r1_pureInitial}) can only be satisfied if $M=d^N$, which implies the rank-one POVM measurement is 
	completely orthogonal. If there are constraints which are not independent, the overcomplete rank-one indecomposable POVM 
	should be employed for optimal information extraction~\cite{Holevo1973,Davies1978information,Clarke2001}.

	Next, we present the derivation of the necessary conditions (\ref{maxInfoExtrac_nC_r1POVM}) and 
	(\ref{maxInfoExtrac_nC_r1_pureInitial}) for optimal information extraction. The joint convexity of the relative 
	entropy implies that $H(m:\pmb{\varphi})$ is a convex function of the initial state $\hat{\rho}_0$ and
	the POVM measurement $\{\hat{\pi}_m\}$~\cite{nielsen2010quantum,cover2006elements}.  Hence there exist the pure initial 
	state $\hat{\rho}_0=\ket{\psi_0}\bra{\psi_0}$ and the corresponding indecomposable measurement which together accomplish 
	the optimal information extraction~\cite{Davies1978information,Chiribella2006}. The rank-one POVM measurement obtained via 
	the spectrum decomposition of an indecomposable POVM measurement is also 
	indecomposable~\cite{PARTHASARATHY1999,DAriano2005}.  It is also known that classifying two different outcomes as as one 
	would not improve the information extraction (cf. Lemma 2 of Ref.~\cite{Davies1978information}). Thus if the rank-one POVM
	measurement is obtained via spectrum decomposition of a POVM measurement, the former would outperform the latter.
	Hence there is always a rank-one indecomposable POVM measures ensures maximal $H(m:\pmb{\varphi})$. Further, 
	the rank-one POVM measurement $\{\hat{\pi}_m=\lambda_m\ket{u_m}\bra{u_m}\}$ can 
	be implemented as a completely orthogonal measurement $\{\hat{\Pi}_m=\ket{v_m}\bra{v_m}\}$ on the composite system 
	of the probes and an auxiliary such that $\ket{v_m} = \hat{V}^\dag\ket{u_m}\otimes\ket{a_m}$ and $\ket{a_m}$ belongs to 
	an orthonormal basis of the auxiliary. $\hat{V}$ is an unitary gate acting upon the composite system
	\begin{equation}
		\hat{V}\ket{\psi}\otimes\ket{a_1} = \sum_{m=1}^M\sqrt{\lambda_m}\ket{u_m}\bra{u_m}\otimes\ket{a_m}.
	\end{equation}
	The optimal $H(m:\pmb{\varphi})$ considered in the subsystem of the probes is thus a conditioned optimal in the composite system 
	of the probes and auxiliary. There are two restrictions: a) the auxiliary is initialized in the standard pure state $\ket{a_1}$ and
	does not pass through the parameter channel; b) only completely orthogonal measurements on the composite system of the probes and 
	auxiliary are considered. The set of strategy combinations of pure initial state $\ket{\psi_0}\otimes\ket{a_1}$ and 
	completely orthogonal measurement $\{\hat{\Pi}_m\}$ 
	corresponds to a connected and compact hypersurface with no boundary in an Euclidean space. Hence the ``maximum point" is 
	also a ``critical point". We choose $\ket{\psi_0}$ as a reference initial state of the probes and $\{\ket{m}\bra{m}\}$ as a reference 
	completely orthogonal measurement on the composite system. Our initialization and measurement setup can thus be described by two unitary 
	gates $\hat{U}_\mathrm{I}$ and $\hat{U}_\mathrm{M}$ such that $\ket{\psi}=\hat{U}_\mathrm{I}\ket{\psi_0}$ and 
	$\hat{\Pi}_m=\hat{U}_\mathrm{M}\ket{m}\bra{m}\hat{U}_\mathrm{M}^\dag$. If the strategy is optimal, the variation of the mutual information  
	\begin{equation}
		{\delta}H(m:\pmb{\varphi})
		= -\int{\ud^{n}\pmb{\varphi}}p_{\pmb{\varphi}}
		\sum_{{m}}\ln\left\lbrack{\frac{p_{{m}}}{p({m}|\pmb{\varphi})}}\right\rbrack\delta{p({m}|\pmb{\varphi})}
		\label{mutualInfo_variation}
	\end{equation}
	 will vanish in the first order of the variation of the initialization gate $\hat{U}_\mathrm{I}$ and the measurement 
	 adjustment gate $\hat{U}_\mathrm{M}$.  The variation of the conditional probability $p(m|\pmb{\varphi})$ due to the 
	 variation of the initialization and the completely orthogonal measurement is
	\begin{equation}
		{\delta}p(m|\pmb{\varphi})
		= i\tr\left(\lbrack{\hat{\rho}_{\pmb{\varphi}},\hat{\Pi}_m}\rbrack\delta\hat{G}_\mathrm{M} 
		+\lbrack{(\hat{U}_{\pmb{\varphi}}^{\otimes{N}})^\dag\hat{\Pi}_m
		\hat{U}_{\pmb{\varphi}}^{\otimes{N}},\hat{\rho}_0}\rbrack\delta\hat{G}_\mathrm{I}\right),
		\label{p_m|varphi_variation}
	\end{equation}
	where $\delta\hat{G}_\mathrm{I}{\defeq}i\hat{U}_\mathrm{I}^\dag\delta\hat{U}_\mathrm{I}$ and 
	$\delta\hat{G}_\mathrm{M}{\defeq}i\hat{U}_\mathrm{M}^\dag\delta\hat{U}_\mathrm{M}$ are Hermitian. Notice that the 
	freedom of choosing arbitrary pure state of the probes and completely orthogonal via the unitary gates 
	$\hat{U}_\mathrm{I}$ and $\hat{U}_\mathrm{M}$ is equivalent to the arbitrariness of $\delta\hat{G}_\mathrm{I}$ and 
	$\delta\hat{G}_\mathrm{M}$.
	Since the initialization gate $\hat{U}_\mathrm{I}$ acts only on the probes while the measurement adjustment gate
	$\hat{U}_\mathrm{M}$ acts on the composite of the probes and the auxiliary, $\delta\hat{G}_\mathrm{I}$ should only act on 
	the probes and $\delta\hat{G}_\mathrm{M}$ acts on the composite system.
	By substituting (\ref{p_m|varphi_variation}) into (\ref{mutualInfo_variation}), we can obtain 
	\begin{equation}
			 \int{\ud^{n}\pmb{\varphi}}p_{\pmb{\varphi}}\sum_{m}\ln\left\lbrack{\frac{p_{{m}}}{p({m}|\pmb{\varphi})}
			 }\right\rbrack 
			\left\lbrack{\hat{\rho}_{\pmb{\varphi}}\otimes\ket{a_0}\bra{a_0},\ket{v_m}\bra{v_m}}\right\rbrack
			=0,
			\label{extreme_infoExtraction_measurementC_COM_composite}
		\end{equation}
		and
		\begin{equation}
			\int{\ud^{n}\pmb{\varphi}}p_{\pmb{\varphi}}\sum_{m}\ln\left\lbrack{\frac{p_{{m}}}{p({m}|\pmb{\varphi})}}\right\rbrack 
			\left(\hat{U}_{\pmb{\varphi}}^{\otimes{N}}\right)^\dag\left\lbrack{\hat{\pi}_m,
			\hat{\rho}_{\pmb{\varphi}}}\right\rbrack\hat{U}_{\pmb{\varphi}}^{\otimes{N}}
			=0.
			\label{extreme_infoExtraction_measurementC_pureInitial_probes}
		\end{equation}
		Expand (\ref{extreme_infoExtraction_measurementC_COM_composite}) to different blocks by the orthogonal projections 
		$\openone\otimes\ket{a_m}\bra{a_m}$ with $m=1,\ldots,M$. Then it is clear that (\ref{extreme_infoExtraction_measurementC_COM_composite}) 
		is equivalent to (\ref{maxInfoExtrac_nC_r1POVM}). If we exapnd (\ref{extreme_infoExtraction_measurementC_pureInitial_probes}) in the orthnormal basis 
		which contains $\ket{\psi_0}$, we can realize immediately that (\ref{extreme_infoExtraction_measurementC_pureInitial_probes}) is equivalent to
		(\ref{maxInfoExtrac_nC_r1_pureInitial}).

\emph{Connection with the conventional variance-covariance picture.}---
	Quantum metrology has conventionally been studied using the covariance matrix for assessing a estimation which reduces to the 
	variance in the single-parameter estimation 
	case~\cite{giovannetti2004quantum-enhanced,giovannetti2006quantum,giovannetti2011advances,braunstein1994statistical}
	\begin{equation} 
		\Delta_\varphi 
		\defeq \braket{\left(\varphi_\est\left/\left|\partial_\varphi{\Saverage{\varphi_\est}}\right|\right.
		-\varphi\right)^2}_\mathrm{av}
	\end{equation}
	where  $\varphi_\est$ is an estimation to the parameter $\varphi$. The optimal strategy in the conventional variance-covariance 
	picture~\cite{Humphreys2013,Baumgratz2016,yuan2016sequential,Pezze2017} is not exactly the same as its counterpart 
	(\ref{maxInfoExtrac_nC_r1POVM}) and (\ref{maxInfoExtrac_nC_r1_pureInitial}) in the information-theoretic picture. 
	There is a intimate relation between the two pictures of quantum metrology in the single-parameter estimation scenario.	
	In a single-parameter estimation, $\Delta_\varphi$ is bounded from below by 
	\begin{equation}
		\Delta_\varphi 
		\ge 1/{\sqrt{s}WN^\alpha},
		\label{rmes_bounds}
	\end{equation}
	where $W$ is the width of the spectrum of the probe Hamiltonian $\hat{\mathcal{H}}$ such that 
	$\hat{U}_\varphi=e^{-i\varphi\hat{\mathcal{H}}}$. The estimation $\varphi_\est$ is extracted from $s$ experiment data points. 
	If an entangled initial state is employed in the parallel strategy or one uses a sequential strategy, the Heisenberg limit 
	$\alpha=1$ can be obtained. Otherwise, only the standard quantum limit $\alpha=1/2$ is available. In the information-theoretic 
	picture, the standard quantum limit of the mutual information $H(m:\varphi)$ is $\frac{1}{2}\ln{N}$ and its Heisenberg is limit 
	$\ln{N}$~\cite{Hassani2017}. It can be shown that the mutual information between the estimation $\varphi_\est$ and the parameter
	$\varphi$ is asymptotically upper bounded as
	\begin{equation}
		H(\varphi_\est:\varphi)
		\le \ln(\sqrt{s^{1+\epsilon}}WN^{\alpha}) +  H(\varphi)
		\label{info_theo_bounds_2nd}
	\end{equation}
	with $\epsilon\ll1$, if the estimation error $\Delta_\varphi$ is lower bounded as (\ref{rmes_bounds}). Further, if 
	(\ref{rmes_bounds}) is asymptotically saturated so would be (\ref{info_theo_bounds_2nd}). As a consequence, any estimation 
	strategy achieves the asymptotic Heisenberg limit (standard quantum limit) in the variance description (\ref{rmes_bounds}) 
	would also obtain the information-theoretic Heisenberg limit (standard quantum limit). It confirms the general connection
	between information-theoretic picture and the variance-covariance description of quantum metrology .

	Next, we will prove that the information-theoretic bound (\ref{info_theo_bounds_2nd}) can be obtained from (\ref{rmes_bounds})
	in the asymptotic sense. Given $r$ times of repetition of the aforementioned estimation which produces $s$ data points, one can
	obtain a final estimation
	\begin{equation}
		\tilde{\varphi}_\est = \frac{1}{r}\sum_{j=1}^r\varphi_\est^{(j)},
	\end{equation}
	by averaging the $r$ estimations $\varphi_\est$. We use the superscript $(j)$ to indicate from which group of data is the 
	estimation $\varphi_\est$ made. For a reasonable 
	estimator one should expect the average of the estimation $\Saverage{\varphi_\est}$ asymptotically approaches the parameter 
	$\varphi$ and $\partial{\Saverage{\varphi_\est}}/\partial{\varphi}$ approaches unity, when $N$ grows 
	bigger~\cite{helstrom1976quantum,giovannetti2011advances,Demkowicz-Dobrzanski2015}.
	In the case of large  $N$, we should have $\Saverage{\tilde{\varphi}_\est} = \varphi$ with 
	$\sigma_\varphi = \Delta_\varphi/\sqrt{r}$ being the standard deviation of $\tilde{\varphi}_\est$. We presume the number of 
	repetition $r$ being sufficiently large and $\varphi_\est^{(j)}$s are mutually independent. Then the central limit theorem
	indicats that the probability of $\tilde{\varphi}_\est$ conditional on $\varphi$ should be Gaussian~\cite{feller1971introduction}
	\begin{equation}
		p(\tilde{\varphi}_\est|\varphi) 
		\simeq \exp\left\lbrack{-{\left(\tilde{\varphi}_\est-\varphi\right)^2}/{2\sigma_\varphi^2}}\right\rbrack
					\Big/\sqrt{2\pi\sigma_\varphi^2}.
	\end{equation}
	Given the prior probability distribution $q_{\varphi}$ of the parameter $\varphi$, we can calculate the information of 
	$\tilde{\varphi}_\est$ conditioned on $\varphi$
	\begin{eqnarray}
		H(\tilde{\varphi}_\est|\varphi)
		&=& -\int\ud\varphi\ud\tilde{\varphi}_{\est}q_{\varphi}p(\tilde{\varphi}_\est|\varphi)\ln{p}(\tilde{\varphi}_\est|\varphi)
		\nonumber\\
		&=& \frac{1}{2}\ln\left(2\pi{e}/r\right) 
			+  \int\ud{\varphi}q_{\varphi}\ln\Delta_\varphi.
		\label{conditional_information_varphi_est_varphi}
	\end{eqnarray}
	With $\sigma_\varphi$ being small enough, the conditional probability $p(\tilde{\varphi}_\est|\varphi)$ would be mainly 
	concentrated within a small area of $|\tilde{\varphi}_\est-\varphi|<k\sigma_\varphi$ such that it can be approximated by
	a delta function $\delta(\tilde{\varphi}_\est-\varphi)$~\cite{arfken2013mathematical}. We can choose 
	$k=\frac{\sigma_\mathrm{max}}{\sigma_\mathrm{min}}
	   \sqrt{\ln(6/\sigma_\mathrm{max})}$ with $\sigma_\mathrm{min}\defeq\min_{\varphi}\sigma_\varphi$ and
	$\sigma_\mathrm{max}\defeq\max_{\varphi}\sigma_\varphi$ such that 
	$k\sigma_\varphi$ is still very small and approaches $0$ very fast as $\sigma_\mathrm{max}\to0$. As a result the 
	Shannon information of $\tilde{\varphi}_\est$ would approach to that of $\varphi$ 
	\begin{equation}
		H(\tilde{\varphi}_\est) 
		= \left\lbrack{1+\mathcal{O}\left({1}/{r}\right)}\right\rbrack{H}(\varphi)+o\left({\sigma_\mathrm{max}}/\sqrt{r}\right),
		\label{varphi_est_shannonInfo}
	\end{equation}
	provided that $q_{\varphi}$ is fairly smooth, namely its first derivative $q^{(1)}(\varphi)$ and the second derivative 
	$q^{(2)}(\varphi)$ with respect to $\varphi$ exist and are finite. Here we  presume the complexity of estimating $\varphi$ of
	different values is of the same order, namely $\sigma_\varphi$s  are of the same order for different values of $\varphi$. For the 
	convenience of mathematical analysis we also assume the first derivative $\sigma_\varphi^{(1)}$ and the second derivative 
	$\sigma_\varphi^{(2)}$ of $\sigma_\varphi$ with respect to $\varphi$ exist and are finite. $\mathcal{O}\left({1}/{r}\right)$ 
	means terms of the same as or higher order than $1/r$ and $o\left({\sigma_\mathrm{max}}/\sqrt{r}\right)$ denotes the terms much 
	smaller than ${\sigma_\mathrm{max}}/\sqrt{r}$. One can obtain the mutual information 
	$H(\tilde{\varphi}_\est:\varphi)=H(\tilde{\varphi}_\est)-H(\tilde{\varphi}_\est|\varphi)$ between $\tilde{\varphi}_\est$ and 
	$\varphi$ by subtracting (\ref{conditional_information_varphi_est_varphi}) from (\ref{varphi_est_shannonInfo}). Then we 
	obtain the upper bound of the mutual information
	\begin{equation}
		H(\tilde{\varphi}_\est:\varphi)
		\le \ln\left(\sqrt{rs}WN^{\alpha}\right) + H(\varphi).
		\label{info_theo_bounds}
	\end{equation}
	from (\ref{rmes_bounds}). We can show that the contribution from $r$ in (\ref{info_theo_bounds}) is superficial. First of all, 
	the mere effect of the $r$ repetition of estimation is 
	increasing the amount of data from $s$ to $sr$. Since $\tilde{\varphi}_\est$ is the average of $r$ estimations each of which is 
	constructed from $s$ data point, $\tilde{\varphi}_\est$ cannot be a better estimation than a general estimation 
	$\varphi_\est$ evaluated directly from the $sr$ experiment data. On the other hand, the scaling behavior of such a general  
	estimation with respect to  $sr$ in the variance picture (\ref{rmes_bounds}) can never be better than $1/\sqrt{sr}$. It means that 
	the two data processing algorithms are equally good asymptotically when both $r$ and $s$ are sufficiently large. Therefore, $r$ 
	can be incorporated to $s$ while $\tilde{\varphi}_\est$ can be replace by an estimation $\varphi_\est$ constructed from $sr$ 
	data points. Secondly, by using the jointly convex property of the mutual information which is a result of the joint convexity
	of the relative entropy~\cite{nielsen2010quantum,cover2006elements}, we can show that
	$H(\varphi_\est:\varphi){\le}H(\tilde{\varphi}_\est:\varphi)$. One can then chose $r=\ln{s}$ and 
	(\ref{info_theo_bounds}) is equivalent to (\ref{info_theo_bounds_2nd}) with 
	$\epsilon=\ln\left(\ln{s}\right)/\ln{s}\ll1$ given that $s$ is big.

\begin{figure*}[th!]
		\subfigure{\includegraphics[width=1.\textwidth]{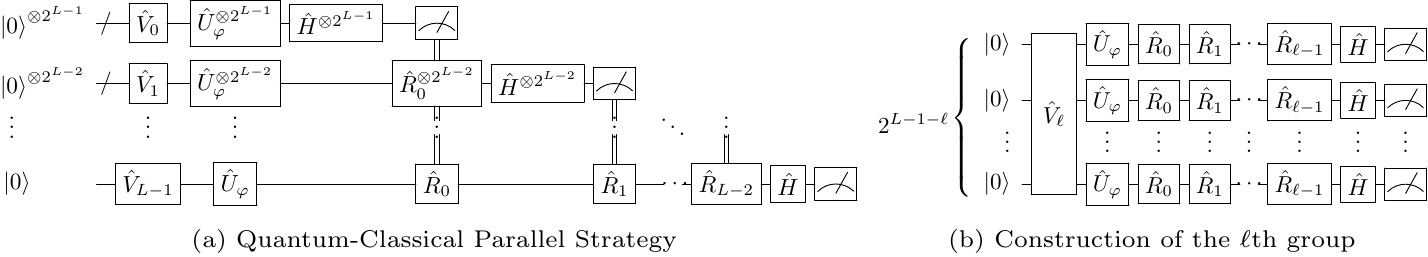}}
		\caption{Quantum-Classical Parallel Strategy.
		$\hat{R}_{\ell}{\defeq}\ket{0}\bra{0}+\exp(i2{\pi}/2^{L-\ell})\ket{1}\bra{1}+\hat{\Pi}$ and
		$\hat{H}{\defeq}\frac{1}{\sqrt{2}}(\ket{0}\bra{0}+\ket{0}\bra{1}+\ket{1}\bra{0}-\ket{1}\bra{1})+\hat{\Pi}$ where
		$\hat{\Pi}\defeq\openone-\ket{0}\bra{0}-\ket{1}\bra{1}$ is a projection onto the subspace perpendicular to the subspace 
		spanned by $\ket{0}$ and $\ket{1}$. Given the probes being two-level systems, $\hat{R}_{\ell}$ are  rotations along the $z$
		axis, $\hat{H}$ is the so-called Hadamard  gate and $\hat{\Pi}$ vanishes. All the rotations $\hat{R}_\ell$ are carried out 
		only when the classical bit transmitted is $1$. The gates implemented after $\hat{U}_\varphi^{\otimes2^{L-1-\ell}}$ plus
		the final separable measurement on the energy eigenbasis in each line of (a) is equivalent to making some separable local
		measurement based on the measurement results of the former groups.
		}
		\label{qc_parallelStrategy}
	\end{figure*}

\emph{Information-theoretic Heisenberg limit of the Quantum-Classical parallel strategy.}--- We know that the entangled measurement
	is not necessary to achieve the conventional Heisenberg limit (\ref{rmes_bounds})~\cite{giovannetti2006quantum}. A direct
	consequence of (\ref{info_theo_bounds_2nd}) is that entangled measurement is not necessary to accomplish the 
	information-theoretic Heisenberg limit either. We can verify this explicitly in a general Quantum-Classical parallel strategy of 
	quantum metrology where the probes are initialized in a GHZ state while a separate measurement is employed at the 
	end~\cite{giovannetti2006quantum}. The scheme employed here (cf. Fig.~\ref{qc_parallelStrategy})
	is similar to the Parallel QPEP presented in Ref.~\cite{Hassani2017} except for several differences. 
	The first difference is the 
	that Hamiltonian $\hat{\mathcal{H}}$ is of arbitrary form as long as its is known and the dimension $d$ of the  Hilbert space of
	a single probe is finite. Secondly, we employ neither CNOT gate nor any other kind of multi-partite gate at the measurement 
	stage. Lastly, 
	we replace the inverse Quantum Fourier Transformation with a modified Semiclassical Fourier 
	Transformation~\cite{griffiths1996semiclassical}. The $N$ probes are divided into $L=\ln(N+1)$ groups and label them with 
	$0,1,\ldots,L-1$. The $\ell$th group has $2^{L-1-\ell}$
	probes as shown in Fig.~\ref{qc_parallelStrategy}(b). The multi-probe gates $\hat{V}_\ell$ would initialize the $2^{L-1-\ell}$
	probes to the GHZ state  which would be transformed by the parameter-imprinting channel $\hat{U}_\varphi^{\otimes2^{L-1-\ell}}$ 
	with $\hat{U}_\varphi=e^{-\varphi\hat{\mathcal{H}}}$ to
	\begin{equation}
		\lbrack\ket{0}^{\otimes2^{L-1-\ell}}+e^{-i2^{L-1-\ell}W\varphi}\ket{1}^{\otimes2^{L-1-\ell}}\rbrack/\sqrt{2}.
		\label{lth_row_info-imprinting}
	\end{equation}
	Here $\ket{0}$ is the ground state of a probe while $\ket{1}$ is its highest excited state. Notice that $W$ is the width of the  
	spectrum of $\hat{\mathcal{H}}$ and we ignored the unimportant global phase in (\ref{lth_row_info-imprinting}). The effect of the
	Semiclassical Fourier Transformation shown in Fig.~\ref{qc_parallelStrategy}(a) plus separable local
	measurement of the probe Hamiltonian is equivalent to the following procedure: Firstly we deploy a separate measurement of 
	$\hat{X}^{\otimes2^{L-1}}$ on the $0$th group with $\hat{X}\defeq\ket{0}\bra{1}+\ket{0}\bra{1}$.  If the measurement tells us 
	that the number of probes having the physical quantity $\hat{X}$ of value $-1$ in the $0$th group is odd, we record the 
	measurement result by $m_0=1$, otherwise $m_0=0$. We mark that the eigenvalue $-1$ of $\hat{X}$ corresponds to 
	the signal $1$ in the quantum circuit shown of Fig.~\ref{qc_parallelStrategy} which signifies the highest level of the probe 
	Hamiltonian $\hat{\mathcal{H}}$. We move forward by measuring $\hat{A}_\ell^{\otimes2^{L-1-\ell}}$ on the 
	$\ell$th group of $2^{L-1-\ell}$ probes according to the measurement results $m_0,\ldots,m_{\ell-1}$ of former groups by 
	carefully choosing
	\begin{equation}
		\hat{A}_{\ell}{\defeq} \hat{X}\cos\left(2\pi\sum_{j=0}^{\ell-1}\frac{{m_j}2^j}{2^{L}}\right)
								-\hat{Y}\sin\left(2\pi\sum_{j=0}^{\ell-1}\frac{{m_j}2^j}{2^{L}}\right),
	\end{equation}
	where $\hat{Y}{\defeq}i(-\ket{0}\bra{1}+\ket{1}\bra{0})$. As one may have realized that  $\hat{X}$ ($\hat{Y}$) is the spin 
	along the $x$ ($y$) direction in the spin-$1/2$ system. Similarly, we
	record $m_\ell=0,1$ according to whether we obtain even or odd number of probes having the physical quantity $\hat{A}_\ell$ of 
	value $-1$ in the $\ell$th group. The conditional probability of obtaining the measurement result $m=\sum_{\ell=0}^{L-1}m_\ell2^\ell$ 
	conditioned on the parameter $\varphi$ is
	\begin{equation}
		p(m|\varphi) = \frac{\sin^2\left\lbrack{(N+1)W\varphi/2}\right\rbrack}
							{(N+1)^2\sin^2\left\{\left\lbrack{W\varphi-2{\pi}m/(N+1)}\right\rbrack/2\right\}}.
	\end{equation}
	With the energy spectrum width $W$ of $\hat{\mathcal{H}}$ set to $1$, we obtain the mutual information between $m$ and $\varphi$
	\begin{equation}
		H(m:\varphi) \simeq \ln{N} - 1.22,
		\label{qc_mutualinformation}
	\end{equation}
	given an uniform prior probability distribution in the interval $0\le\varphi\le2\pi$. It is the exactly the same as the result 
	presented for QPEP in Ref.~\cite{Hassani2017}.  Note that though we do not employ any 
	entangled measurement, the measurement employed here is made possible with the assistant of classical 
	communication~\cite{nielsen2010quantum}.

\emph{The Width of energy spectrum and the Hilbert space dimension.}---Just as in the variance-covariance picture (\ref{rmes_bounds}),
	it is the energy spectrum width not the dimension of the Hilbert space of the probe that enters the standard quantum limit 
	$\alpha=1/2$ and the Heisenberg limit $\alpha=1$ in the information-theoretic picture (\ref{info_theo_bounds_2nd}). This can also
	be seen explicitly in (\ref{qc_mutualinformation}) of a general quantum-classical parallel strategy of estimation. The dimension
	$d$ of the probe Hilbert space can be involved in some special cases. One typical example is the case wheres the probe energy 
	levels are equally separated $\hat{\mathcal{H}}=-\sum_{k=0}^{d-1}2k\pi$ as presented in Ref.~\cite{Hassani2017}. The probe energy
	spectrum width and Hilbert space dimension would then be equivalent $W=2\pi(d-1)$. In such as case, the effect of the probe 
	Hilbert space dimension $d$ would manifest itself as $\ln(d-1)$ for both the standard quantum limit and the Heisenberg limit.

\emph{Conclusion.}---We have investigated the optimal information extraction strategy in quantum metrology. The general method to
	obtain such optimal strategy is outlined via (\ref{maxInfoExtrac_nC_r1POVM}) and (\ref{maxInfoExtrac_nC_r1_pureInitial}). Our result can be
	generalized to information extraction task such as the discrimination of quantum states and quantum channels.
	We have further established a direct connection between the conventional variance-covariance picture and the 
	information-theoretic picture of quantum metrology in the single-parameter estimation scenario. This connection indicates that 
	the usefulness of entanglement at different stages of a metrology task is the same in both pictures. We explicitly illustrate 
	this by showing a slightly modified  Quantum-Classical Parallel strategy which can achieve the Heisenberg limit in both pictures.
	Our results also indicate that it is the energy spectrum width rather than the dimension of the probe which plays an nontrivial role in quantum 
	metrology in the information-theoretic pictures.
\begin{acknowledgments}
	We would like to than M. J. W. Hall for invaluable comment on this work. This work was supported by Ministry of Science 
	and Technology of China (Grants No. 2016YFA0302104 and 2016YFA0300600), National Natural Science Foundation of China 
	(Grant Nos. 91536108 and 11774406) and Chinese Academy of Sciences (XDPB08-3).
\end{acknowledgments}
\bibliography{Bibliography.bib}
\end{document}